\PassOptionsToPackage{unicode=true}{hyperref} 
\PassOptionsToPackage{hyphens}{url}
\documentclass[
  twocolumn]{article}
\usepackage{lmodern}
\usepackage{amssymb,amsmath}
\usepackage{ifxetex,ifluatex}
\ifnum 0\ifxetex 1\fi\ifluatex 1\fi=0 
  \usepackage[T1]{fontenc}
  \usepackage[utf8]{inputenc}
  \usepackage{textcomp} 
\else 
  \usepackage{unicode-math}
  \defaultfontfeatures{Scale=MatchLowercase}
  \defaultfontfeatures[\rmfamily]{Ligatures=TeX,Scale=1}
\fi
\IfFileExists{upquote.sty}{\usepackage{upquote}}{}
\IfFileExists{microtype.sty}{
  \usepackage[]{microtype}
  \UseMicrotypeSet[protrusion]{basicmath} 
}{}
\makeatletter
\@ifundefined{KOMAClassName}{
  \IfFileExists{parskip.sty}{%
    \usepackage{parskip}
  }{
    \setlength{\parindent}{0pt}
    \setlength{\parskip}{6pt plus 2pt minus 1pt}}
}{
  \KOMAoptions{parskip=half}}
\makeatother
\usepackage{xcolor}
\IfFileExists{xurl.sty}{\usepackage{xurl}}{} 
\IfFileExists{bookmark.sty}{\usepackage{bookmark}}{\usepackage{hyperref}}
\hypersetup{
  pdftitle={A bound on the Treatment Effect Heterogeneity using estimates on the Variability Ratio},
  pdfauthor={Alexander Volkmann},
  pdfborder={0 0 0},
  breaklinks=true}
\usepackage[margin=1in]{geometry}
\usepackage{color}
\usepackage{fancyvrb}

\DefineVerbatimEnvironment{Highlighting}{Verbatim}{commandchars=\\\{\}}
\usepackage{framed}
\definecolor{shadecolor}{RGB}{248,248,248}
\newenvironment{Shaded}{\begin{snugshade}}{\end{snugshade}}

\newcommand{\BuiltInTok}[1]{#1}

\newcommand{\CommentTok}[1]{\textcolor[rgb]{0.56,0.35,0.01}{\textit{#1}}}

\newcommand{\ControlFlowTok}[1]{\textcolor[rgb]{0.13,0.29,0.53}{\textbf{#1}}}

\newcommand{\DecValTok}[1]{\textcolor[rgb]{0.00,0.00,0.81}{#1}}

\newcommand{\FloatTok}[1]{\textcolor[rgb]{0.00,0.00,0.81}{#1}}

\newcommand{\ImportTok}[1]{#1}

\newcommand{\KeywordTok}[1]{\textcolor[rgb]{0.13,0.29,0.53}{\textbf{#1}}}
\newcommand{\NormalTok}[1]{#1}
\newcommand{\OperatorTok}[1]{\textcolor[rgb]{0.81,0.36,0.00}{\textbf{#1}}}

\newcommand{\StringTok}[1]{\textcolor[rgb]{0.31,0.60,0.02}{#1}}
\newcommand{\VariableTok}[1]{\textcolor[rgb]{0.00,0.00,0.00}{#1}}

\usepackage{graphicx,grffile}
\makeatletter
\def\maxwidth{\ifdim\Gin@nat@width>\linewidth\linewidth\else\Gin@nat@width\fi}
\def\maxheight{\ifdim\Gin@nat@height>\textheight\textheight\else\Gin@nat@height\fi}
\makeatother
\setkeys{Gin}{width=\maxwidth,height=\maxheight,keepaspectratio}
\ifx\paragraph\undefined\else
  \let\oldparagraph\paragraph
  \renewcommand{\paragraph}[1]{\oldparagraph{#1}\mbox{}}
\fi
\ifx\subparagraph\undefined\else
  \let\oldsubparagraph\subparagraph
  \renewcommand{\subparagraph}[1]{\oldsubparagraph{#1}\mbox{}}
\fi

\makeatletter
\def\fps@figure{htbp}
\makeatother

\title{On the Relationship between Treatment Effect Heterogeneity and the Variability Ratio Effect Size Statistic}
\author{Alexander Volkmann\footnote{Marienburger Str. 21, 10405 Berlin,
  \href{mailto:alexv@gmx.de}{\nolinkurl{alexv@gmx.de}}}}
\date{}

\begin{document}
\maketitle

\hypertarget{abstract}{%
\section{Abstract}\label{abstract}}

Recently, the variability ratio (VR) effect size statistic has been used with
increasing frequency in the study of differences in variation of a
measured variable between two study populations. More specifically, the
VR effect size statistic allows for the detection of treatment effect
heterogeneity (TEH) of medical interventions. While a VR that is
different from \(1\) is widely acknowledged to implicate a treatment
effect heterogeneity (TEH) the exact relationship between those two
quantities has not been discussed in detail thus far.

In this note we derive a precise connection between TEH and VR. In
particular, we derive precise upper and lower bounds on the TEH in terms
of VR. Moreover, we provide an exemplary simulation for which VR is
equal to \(1\) and there exist TEH.

Our result has implications for the interpretation of VR effect size
estimates regarding its connection to treatment effect heterogeneity of
(medical) interventions.

\newcommand*{\keywords}[1]{\textbf{\textit{Keywords ---}} #1}
\footnotetext{\textbf{Abbreviations:} VR, variability ratio; TEH, treatment effect heterogeneity}
\textbf{\textit{Keywords ---}} variability ratio, treatment effect heterogeneity, causal inference

\section{Introduction}\label{sec1}

Recently, the variability ratio (VR) effect size statistic (defined in Hedges
and Nowell (1995), and proposed in Nakagawa et al. (2015)) has been used
in various meta-analyses to investigate the difference of the total
amount of variance present between two experimental groups.

Senior et al. (2016) compared the effects of two dietary interventions
on variability in weight. Winkelbeiner et al. (2019), McCutcheon et al.
(2019) and Mizuno et al. (2020) studied the variability of antipsychotic
drug response in schizophrenia. Plöderl and Hengartner (2019), Maslej et
al. (2020) and Volkmann, Volkmann, and Mueller (2020) investigated the
variability of antidepressants' response in depression.

Other recent works that have used the VR effect size statistic in order to
compare the variability between two groups include Brugger and Howes
(2017), Pillinger et al. (2019), Rogdaki et al. (2020) as well as
Brugger et al. (2020).

While a VR that is different from \(1\) is widely acknowledged to
implicate a variation in treatment effect the exact relationship between
those two quantities has not discussed in detail thus far.

The purpose of this note is to derive a precise connection between the
TEH and the VR. Moreover, we provide an exemplary simulation for which
VR is equal to \(1\), yet there exist substantial TEH.

\hypertarget{methods-individual-treatment-effect}{%
\section{Methods -- Individual treatment
effect}\label{methods-individual-treatment-effect}}

The treatment effect of individual \(i\) with respect to a given
(medical) treatment is defined as (cf.~e.g.~Morgan and Winship (2015)),

\[\delta_i = Y_{i}^1 - Y_{i}^0,\]

where \(Y_{i}^a\) denotes the potential outcome of individual \(i\) with
respect to treatment \(a \in \{0,1\}\). The potential outcome \(Y_i^a\)
represents the outcome that the individual \(i\) \emph{would have} had,
had the individual \(i\) received the treatment \(a\), irrespective of
the \emph{factual} treatment, the treatment that actually \emph{was}
received by the individual. The \emph{fundamental problem of causal
inference} (see Holland (1986)), states that we can never observe both
\(Y_{i}^1\) and \(Y_{i}^0\) at the same time. Moreover, the notation
reflects the fact that we make the Stable Unit Treatment Value
Assumption (SUTVA) assumption, which allows us to write the potential
outcomes for an individual \(i\) as a function of the individual's own
treatment assignment alone, rather than the treatment assignment of all
individuals in the population. Mathematically,
\(Y_i^{a_1,...,a_i, ...} \equiv Y_i^{a_i}\). This assumption is not
always valid. In case the treatment refers to a vaccination against a
disease the assumptions is violated, since the effect of a potential
vaccination of an individual \(i\) depends on whether others have been
vaccinated.

In the following we consider the so-called super-population perspective,
in which we view our sample \(Y_{i}^0, Y_{i}^1\) of size \(N\) as a
random sample from an infinite population. More formally, we consider
\(Y_i^a\) to be realizations of a population level random variable
\(Y^a\). Similarly, \(\delta_i\) are realizations of a population level
random variable \(\delta\).

In summary, there are two levels of randomness that we consider.
Firstly, we randomly draw individuals \(i=1, ..., N\) from an infinite
population, each individual having an associated pair \((Y_i^0, Y_i^1)\)
of potential outcomes. Secondly, a predetermined number
(\(0 < N_1 < N\)) of individuals get randomly assigned to treatment
(\(a=1\)), the remaining (\(N_0=N-N_1\)) individuals get assigned to the
control (\(a=0\)) group, and the associated outcomes \(Y_i=Y_i^a\) are
observed. The uncertainty in any type of estimate that results from
these two levels of randomness may be referred to as sampling-based and
design-based uncertainty, respectively (see Abadie et al. (2020)).

Notably, there is an additional layer of randomness that might be
considered. We could assume that for each individual \(i\), \(Y_{i}^a\)
are random variables rather than mere realizations of a population level
random variable \(Y^a\). In this case one might refer to \(\delta_i\) as
the random treatment effect of the individual \(i\). Realizations
\(y_i^a\) of \(Y_{i}^a\) could be viewed as draws from a
``\emph{metaphorical population}, comprising the possible eventualities
that might have occurred but mainly didn't'' (see Spiegelhalter (2019)).
Alternatively, the source for this part of the randomness could be a
model for measurement error, epistemological uncertainty, or could
reflect the assumption that the individual \(i\) has a \emph{truly}
random outcome.

\hypertarget{treatment-effect-heterogeneity}{%
\subsection{Treatment effect
heterogeneity}\label{treatment-effect-heterogeneity}}

Treatment effect heterogeneity (TEH) is present if not all individuals
have the same individual treatment effect, i.e.~\(\delta_i\) is not the
same for all individuals. In the language of the super-population
perspective this means that \(Var(\delta) > 0\). Of course, in any
practical situation it is impossible \emph{not} to have TEH. The more
interesting question to study is the question about the degree or the
magnitude of TEH. It can be quantified by estimating the magnitude of
\(Var(\delta)\) or by identifying a relevant (w.r.t. the application at
hand) subgroup \(X\) of the population for which
\(E[\delta|X] - E[\delta]\) is large.

In the case of random individual treatment effect, we need to
distinguish between inter- and intra-individual TEH. The
inter-individual TEH could be defined as the variation between
individuals of the average random individual treatment effect
(\(E[\delta_i]\)).

\hypertarget{variability-ratio}{%
\subsection{Variability Ratio}\label{variability-ratio}}

We define the \emph{variability ratio} \(\nu\) (cf.~Hedges and Nowell
(1995)) as \[\nu = \frac{\sigma_1}{\sigma_0},\] where
\(\sigma_a^2 = Var(Y^a)\), for \(a=0,1\), are the super-population
variances.

In the case of sufficiently large sample sizes \(N_0, N_1\),
sufficiently large values \(\sigma_a\) and (approximately) normally
distributed \(\ln \sigma_a\), an unbiased estimator of \(\ln\nu\) and
its sample variance (cf.~Raudenbush and Bryk (1987); Nakagawa et al.
(2015)) are given by \begin{align*}
\ln VR & = \ln\left(\frac{s_1}{s_0}\right) + \frac{1}{2(N_1-1)} - \frac{1}{2(N_0-1)} \\
s^2_{\ln VR} & = \frac{1}{2(N_1-1)} - \frac{1}{2(N_0-1)}.
\end{align*} Here, \(s_0\) and \(s_1\) denote the finite sample
variances of treatment and control units' observed scores \(Y_i\).

\hypertarget{results-treatment-effect-heterogeneity-and-variability-ratio}{%
\section{Results -- Treatment effect heterogeneity and variability
ratio}\label{results-treatment-effect-heterogeneity-and-variability-ratio}}

In order to derive the main analytical relationship between TEH and the
VR effect size statistic we consider two time points \(t=0\) (baseline) and
\(t=1\) (endpoint). We denote by \(Y_{it}^a\) the potential outcomes of
individual \(i\) at time \(t\) with respect to the treatment \(a\). We
assume that \(Y_{i0}^0 \equiv Y_{i0}^1\), i.e.~there is no effect at
baseline. Then, we may write

\begin{align*}
Y_{i1}^a 
& = Y_{i0}^a + (Y_{i1}^0 - Y_{i0}^a) + a \cdot (Y_{i1}^1 - Y_{i1}^0) \\
& = Y_{i0}^0 + (Y_{i1}^0 - Y_{i0}^0) + a \cdot (Y_{i1}^1 - Y_{i1}^0) \\
& =^{def} \alpha_i + \tau_i + a \cdot \delta_i.
\end{align*}

In words, this means that we can decompose the potential outcome score
\(Y_{i1}^a\) of individual \(i\) at endpoint (\(t=1\)) under treatment
\(a\) as the sum of the following quantities (Figure 1):

\begin{itemize}
\item
  the baseline score (aka the pre-treatment score) of individual $i$
\item
  the temporal change of score (aka the response) of individual $i$
  under control
\item
  the treatment effect of individual $i$ in case $a=1$ and zero
  otherwise.
\end{itemize}

\begin{figure}
\centering
\includegraphics{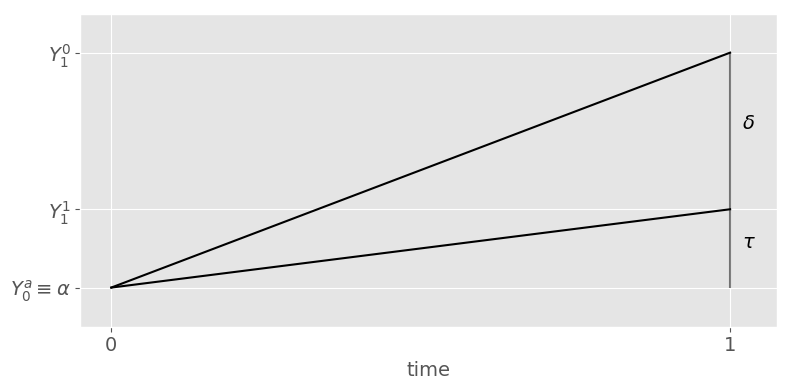}
\caption{Decomposition of the potential outcomes \(Y_1^a\).}
\end{figure}

We then have for the difference in variance in temporal change
\begin{align}\label{raweq}
Var(Y_{1}^1 & - Y_{0}^1) - Var(Y_{1}^0 - Y_{0}^0) \nonumber \\ 
& = Var(\delta) + 2Cov(\tau, \delta) \nonumber \\
& = Var(\delta) + 2\rho \,Var(\tau)^\frac{1}{2}Var(\delta)^\frac{1}{2}.
\end{align}

Here, \(\rho\) denotes the correlation between \(\tau\), the change (or
response) score under control, and \(\delta\), the treatment effect.
Note that this quantity is not observable since it contains information
about the correlation between both potential outcomes \(Y_1^0\) and
\(Y_1^1\).

\hypertarget{compatible-values-of-teh-and-vr}{%
\subsection{Compatible values of TEH and
VR}\label{compatible-values-of-teh-and-vr}}

With equation (\ref{raweq}) at hand, we are ready to derive restrictions
on the standard deviation \(\sigma_{\delta} = Var(\delta)^\frac{1}{2}\)
of the treatment effect \(\delta\).

Let \(\nu\) be the variability ratio defined with respect to the
response variables \(Y^a = Y_{1}^a - Y_{0}^a\). Then, we have that
\begin{align}\label{maineq}
\nu^2 - 1
& = \frac{Var(Y_{1}^1 - Y_{0}^1) - Var(Y_{1}^0 - Y_{0}^0)}{Var(Y_{1}^0 - Y_{0}^0)} \nonumber\\
& =  \frac{ Var(\delta) + 2\rho \,Var(\tau)^\frac{1}{2}Var(\delta)^\frac{1}{2} }{Var(\tau)}\nonumber\\
& =  \frac{ \sigma_{\delta}^2  + 2\rho \,\sigma_\tau \sigma_{\delta}}{\sigma_\tau^2}\nonumber\\
& =  \frac{1}{\sigma_\tau^2}(\sigma_{\delta} + \rho \,\sigma_\tau)^2  - \rho^2 .
\end{align}

Now set \(r := \nu^2 - 1\), then we see that we must have that
\(r\geq -\rho^2\), and we may write
\[|\sigma_{\delta} + \rho\, \sigma_\tau | = \sigma_\tau \sqrt{r+ \rho^2}.\]
Hence, we have the following compatible values for \(\sigma_\delta\) \[
\begin{cases}
\sigma_{\delta} = \sigma_\tau (\sqrt{r+ \rho^2} - \rho)
    ;& r \geq 0 \\
\sigma_{\delta} = 
    \sigma_\tau (\pm\sqrt{r+ \rho^2} - \rho) 
    ; & \rho \leq 0 \;\text{and}\;  -\rho^2 \leq r \leq 0 .
\end{cases}
\]

\begin{figure}
\centering
\includegraphics{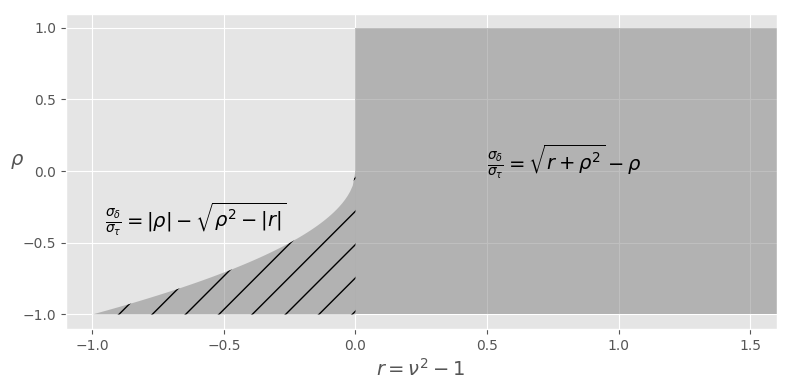}
\caption{Compatible combinations of \(r\) and \(\rho\). Hatched area has
two solutions for \(\sigma_\delta\).}
\end{figure}

It is interesting to observe that for negative values of \(r\) (or
equivalently, \(\nu <1\)) there are two possible solutions for
\(\sigma_\delta\) as visualized in figures (3) and (4).

\begin{figure}
\centering
\includegraphics{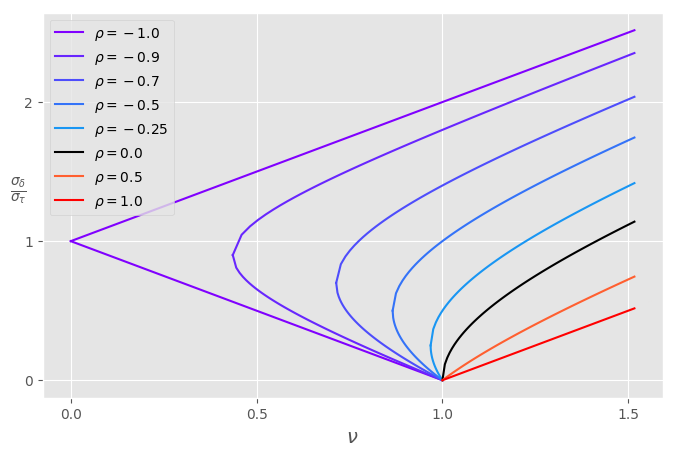}
\caption{Variability ratio \(\nu\) against relative treatment effect
heterogeneity \(\frac{\sigma_\delta}{\sigma_\tau}\) for different levels
of assumed correlation \(\rho\).}
\end{figure}

\begin{figure}
\centering
\includegraphics{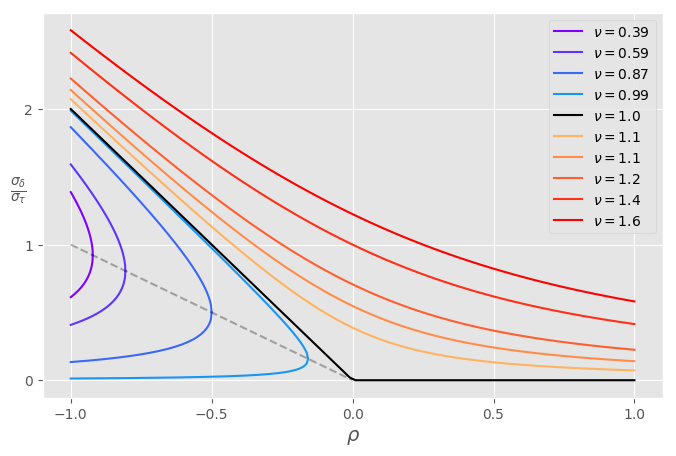}
\caption{Assumed correlation \(\rho\) against relative treatment effect
heterogeneity \(\frac{\sigma_\delta}{\sigma_\tau}\) for different levels
of assumed \(\nu\).}
\end{figure}

Since the quantities \(r=\nu^2-1\) and \(\sigma_\tau\) are estimable
from data, the main equation (\ref{maineq}) implies the following bound
on \(\sigma_\delta\):

\[
|1 - \nu| \leq \frac{\sigma_{\delta}}{\sigma_\tau}  \leq 1 + \nu .
\]

If we are willing to make assumptions on the values of \(\rho\),
e.g.~through domain knowledge, this estimate can be improved. In case
\(\rho=0\) we have that

\[
\sigma_{\delta} = \sigma_\tau \sqrt{\nu^2-1}.
\]

From a Bayesian perspective, we may put a distribution on the values of
\(\rho\) representing our belief about the true value of \(\rho\), which
in turn yields a probability distribution on the values of
\(\sigma_{\delta}\).

\begin{figure}
\centering
\includegraphics{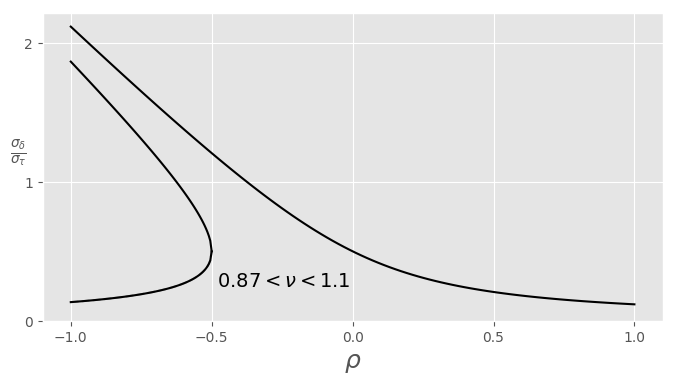}
\caption{Compatible region of \(\rho\) and
\(\frac{\sigma_\delta}{\sigma_\tau}\) assuming exemplary
\(0.87 < \nu < 1.1\).}
\end{figure}

\hypertarget{simulations}{%
\subsection{Simulations}\label{simulations}}

In this section we conduct a simulation that illustrates the
compatibility of a variability ratio of \(1\) and an average treatment
effect of \(0\) with a TEH of \(\sigma_\delta=1\).

We consider the following toy example. Let \(Y^0, Y^1\) be the potential
outcome responses under control and treatment, respectively. We let
\((Y^0, Y^1-Y^0)\) have a distribution with mean \(0\) and covariance
matrix \(\Sigma\) given by \[
\Sigma = 
\begin{pmatrix}
1 & -0.5  \\
-0.5 & 1 
\end{pmatrix}.
\]

The following python code generates potential outcomes of this toy model
for 10000 units:

\footnotesize

\begin{Shaded}
\begin{Highlighting}[]
\ImportTok{import}\NormalTok{ numpy }\ImportTok{as}\NormalTok{ np}
\ImportTok{from}\NormalTok{ numpy.random }\ImportTok{import}\NormalTok{ choice}

\NormalTok{np.random.seed(}\DecValTok{1}\NormalTok{)}

\NormalTok{rho }\OperatorTok{=} \FloatTok{-0.5}
\NormalTok{mu_tau }\OperatorTok{=} \DecValTok{0}
\NormalTok{sigma_tau }\OperatorTok{=} \DecValTok{1}
\NormalTok{mu_delta }\OperatorTok{=} \DecValTok{0}
\NormalTok{sigma_delta }\OperatorTok{=} \DecValTok{1}

\NormalTok{N }\OperatorTok{=} \DecValTok{10000}


\KeywordTok{def}\NormalTok{ draw_potential_outcomes(N):}
\NormalTok{    Sigma }\OperatorTok{=}\NormalTok{ np.array(}
\NormalTok{            [}
\NormalTok{                [}
\NormalTok{                    sigma_tau }\OperatorTok{**} \DecValTok{2}\NormalTok{,}
\NormalTok{                    rho }\OperatorTok{*}\NormalTok{ sigma_delta }\OperatorTok{*}\NormalTok{ sigma_tau}
\NormalTok{                ],}
\NormalTok{                [}
\NormalTok{                    rho }\OperatorTok{*}\NormalTok{ sigma_tau }\OperatorTok{*}\NormalTok{ sigma_delta,}
\NormalTok{                    sigma_delta }\OperatorTok{**} \DecValTok{2}
\NormalTok{                ]}
\NormalTok{            ]}
\NormalTok{        )}
\NormalTok{    u }\OperatorTok{=}\NormalTok{ np.random.multivariate_normal(}
\NormalTok{        (mu_tau, mu_delta), Sigma, size}\OperatorTok{=}\NormalTok{N}
\NormalTok{    )}
\NormalTok{    Y0 }\OperatorTok{=}\NormalTok{ u[:, }\DecValTok{0}\NormalTok{]}
\NormalTok{    Y1 }\OperatorTok{=}\NormalTok{ Y0 }\OperatorTok{+}\NormalTok{ u[:, }\DecValTok{1}\NormalTok{]}
    \ControlFlowTok{return}\NormalTok{ Y0, Y1}
\end{Highlighting}
\end{Shaded}

\normalsize

The python code below conducts 1000 simulations of drawing 10000 units
from the toy model distribution, calculating the empirical standard
deviation of the (unobservable) treatment effect, randomly assigning
them into treatment and control groups, and then calculating the VR
effect size statistic.

\footnotesize

\begin{Shaded}
\begin{Highlighting}[]
\ImportTok{import}\NormalTok{ pandas }\ImportTok{as}\NormalTok{ pd}

\NormalTok{simulations }\OperatorTok{=} \DecValTok{1000}


\KeywordTok{def}\NormalTok{ get_simulation_df(simulations, N):}

\NormalTok{    df }\OperatorTok{=}\NormalTok{ pd.DataFrame()}

    \ControlFlowTok{for}\NormalTok{ i }\KeywordTok{in} \BuiltInTok{range}\NormalTok{(simulations):}
    
\NormalTok{        Y0, Y1 }\OperatorTok{=}\NormalTok{ draw_potential_outcomes(N)}
        \CommentTok{# randomize N units into treatment and control}
\NormalTok{        W }\OperatorTok{=}\NormalTok{ np.array([}\VariableTok{False} \ControlFlowTok{for}\NormalTok{ _ }\KeywordTok{in} \BuiltInTok{range}\NormalTok{(N)])}
\NormalTok{        N1 }\OperatorTok{=} \BuiltInTok{int}\NormalTok{(N }\OperatorTok{/} \DecValTok{2}\NormalTok{)}
\NormalTok{        W[choice(}\BuiltInTok{range}\NormalTok{(N), N1, replace}\OperatorTok{=}\VariableTok{False}\NormalTok{)] }\OperatorTok{=} \VariableTok{True}
        
\NormalTok{        Y1_obs }\OperatorTok{=}\NormalTok{ Y1[W]}
\NormalTok{        Y0_obs }\OperatorTok{=}\NormalTok{ Y0[}\OperatorTok{~}\NormalTok{W]}
        
\NormalTok{        SD_treatment }\OperatorTok{=}\NormalTok{ Y1_obs.std(ddof}\OperatorTok{=}\DecValTok{1}\NormalTok{)}
\NormalTok{        SD_control }\OperatorTok{=}\NormalTok{ Y0_obs.std(ddof}\OperatorTok{=}\DecValTok{1}\NormalTok{)}
\NormalTok{        VR }\OperatorTok{=}\NormalTok{ SD_treatment }\OperatorTok{/}\NormalTok{ SD_control}
        
\NormalTok{        SD_delta }\OperatorTok{=}\NormalTok{ (Y1 }\OperatorTok{-}\NormalTok{ Y0).std()}
    
\NormalTok{        df }\OperatorTok{=}\NormalTok{ df.append(}
\NormalTok{            \{}
                \StringTok{'VR'}\NormalTok{: VR,}
                \StringTok{'SD_delta'}\NormalTok{: SD_delta}
\NormalTok{            \},}
\NormalTok{            ignore_index}\OperatorTok{=}\VariableTok{True}
\NormalTok{        ) }
    \ControlFlowTok{return}\NormalTok{ df}
\end{Highlighting}
\end{Shaded}

\normalsize

The python code above was used to generate the following visualization
of the simulations:

\normalsize

\begin{figure}
\centering
\includegraphics{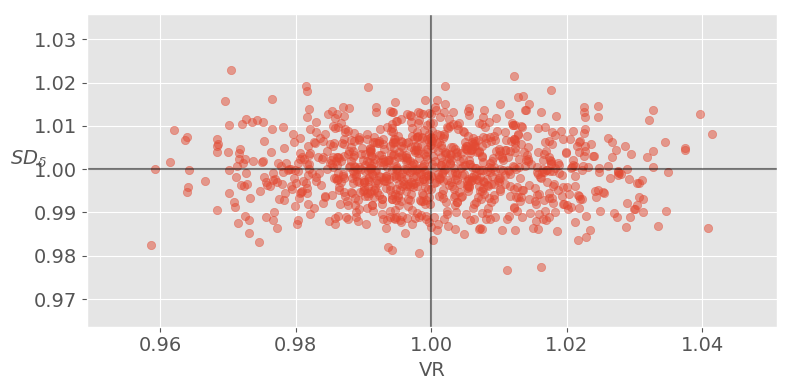}
\caption{Simulated data of 10000 units with \(\nu=1\) and
\(\sigma_\delta=0\).}
\end{figure}

\includegraphics{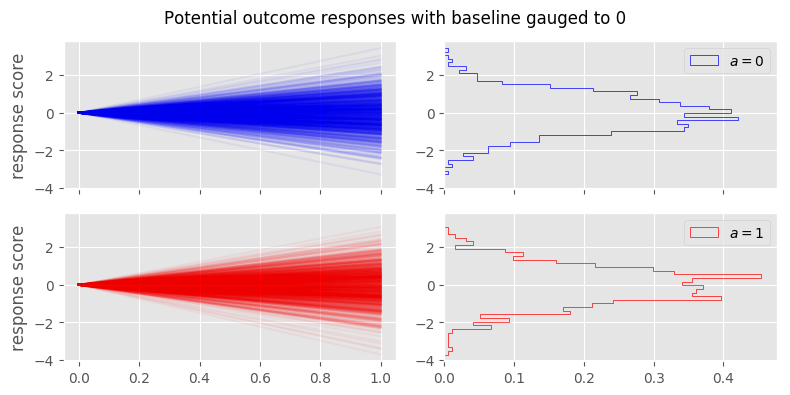}
\includegraphics{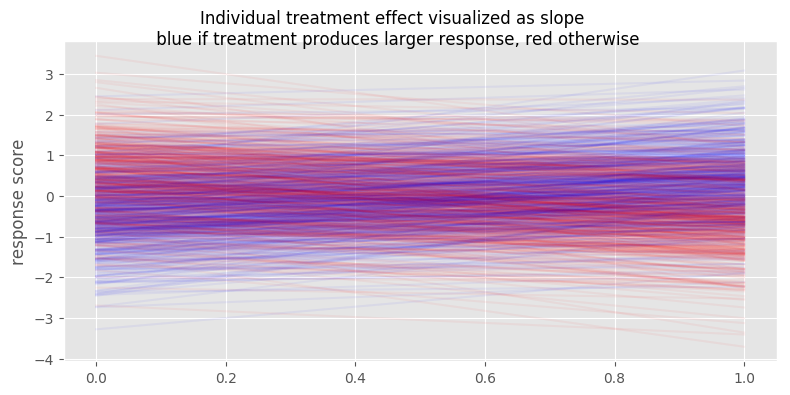}

\hypertarget{discussion-and-conclusions}{%
\section{Discussion and Conclusions}\label{discussion-and-conclusions}}

The variability ratio (VR) effect size statistic (defined in Hedges and
Nowell (1995); and proposed in Nakagawa et al. (2015)) has been used
extensively in order to study treatment effect heterogeneity (TEH) in
clinical studies. While a VR that is different from \(1\) is widely
acknowledged to implicate a treatment effect heterogeneity (TEH) the
exact relationship between those two quantities has not been discussed
in detail thus far.

In this note we derived an analytic expression that connects the VR and
the standard deviation of the treatment effect that includes an
unobservable correlation coefficient. This equation implies precise
upper and lower bounds on the the standard deviation of the treatment
effect in terms of the VR and the standard deviation of the response
under placebo.

In particular, we showed that in case that the variability is equal to
\(1\), the standard deviation of the treatment effect is at most twice
the size of the standard deviation of the response under placebo.
Moreover, if one is willing to make assumptions on the non-negativity of
an unobserved quantity this implies a constant treatment effect. We
illustrated our finding with visualizations and a simulation.

\hypertarget{acknowledgments}{%
\section{Acknowledgments}\label{acknowledgments}}

The inspiration for this work came from joint work with Constantin
Volkmann and Christian Müller (see Volkmann, Volkmann, and Mueller 2020)
on the study of TEH of antidepressants in major depression. I would like
to thank my brother Constantin Volkmann for his interest in this work
and fruitful conversations related to this work.

\hypertarget{references}{%
\section*{References}\label{references}}
\addcontentsline{toc}{section}{References}

\hypertarget{refs}{}
\leavevmode\hypertarget{ref-abadie2020sampling}{}%
Abadie, Alberto, Susan Athey, Guido W Imbens, and Jeffrey M Wooldridge.
2020. ``Sampling-Based Versus Design-Based Uncertainty in Regression
Analysis.'' \emph{Econometrica} 88 (1): 265--96.

\leavevmode\hypertarget{ref-brugger2020heterogeneity}{}%
Brugger, Stefan P, Ilinca Angelescu, Anissa Abi-Dargham, Romina Mizrahi,
Vahid Shahrezaei, and Oliver D Howes. 2020. ``Heterogeneity of Striatal
Dopamine Function in Schizophrenia: Meta-Analysis of Variance.''
\emph{Biological Psychiatry} 87 (3): 215--24.

\leavevmode\hypertarget{ref-brugger2017heterogeneity}{}%
Brugger, Stefan P, and Oliver D Howes. 2017. ``Heterogeneity and
Homogeneity of Regional Brain Structure in Schizophrenia: A
Meta-Analysis.'' \emph{JAMA Psychiatry} 74 (11): 1104--11.

\leavevmode\hypertarget{ref-hedges1995sex}{}%
Hedges, Larry V, and Amy Nowell. 1995. ``Sex Differences in Mental Test
Scores, Variability, and Numbers of High-Scoring Individuals.''
\emph{Science} 269 (5220): 41--45.

\leavevmode\hypertarget{ref-holland1986statistics}{}%
Holland, Paul W. 1986. ``Statistics and Causal Inference.''
\emph{Journal of the American Statistical Association} 81 (396):
945--60.

\leavevmode\hypertarget{ref-maslej2020individual}{}%
Maslej, Marta M, Toshiaki A Furukawa, Andrea Cipriani, Paul W Andrews,
and Benoit H Mulsant. 2020. ``Individual Differences in Response to
Antidepressants: A Meta-Analysis of Placebo-Controlled Randomized
Clinical Trials.'' \emph{JAMA Psychiatry}.

\leavevmode\hypertarget{ref-mccutcheon2019efficacy}{}%
McCutcheon, Robert A, Toby Pillinger, Yuya Mizuno, Adam Montgomery,
Haridha Pandian, Luke Vano, Tiago Reis Marques, and Oliver D Howes.
2019. ``The Efficacy and Heterogeneity of Antipsychotic Response in
Schizophrenia: A Meta-Analysis.'' \emph{Molecular Psychiatry}, 1--11.

\leavevmode\hypertarget{ref-mizuno2020heterogeneity}{}%
Mizuno, Yuya, Robert A McCutcheon, Stefan P Brugger, and Oliver D Howes.
2020. ``Heterogeneity and Efficacy of Antipsychotic Treatment for
Schizophrenia with or Without Treatment Resistance: A Meta-Analysis.''
\emph{Neuropsychopharmacology} 45 (4): 622--31.

\leavevmode\hypertarget{ref-morgan2015counterfactuals}{}%
Morgan, Stephen L, and Christopher Winship. 2015. \emph{Counterfactuals
and Causal Inference}. Cambridge University Press.

\leavevmode\hypertarget{ref-nakagawa2015meta}{}%
Nakagawa, Shinichi, Robert Poulin, Kerrie Mengersen, Klaus Reinhold,
Leif Engqvist, Malgorzata Lagisz, and Alistair M Senior. 2015.
``Meta-Analysis of Variation: Ecological and Evolutionary Applications
and Beyond.'' \emph{Methods in Ecology and Evolution} 6 (2): 143--52.

\leavevmode\hypertarget{ref-pillinger2019meta}{}%
Pillinger, Toby, Emanuele F Osimo, Stefan Brugger, Valeria Mondelli,
Robert A McCutcheon, and Oliver D Howes. 2019. ``A Meta-Analysis of
Immune Parameters, Variability, and Assessment of Modal Distribution in
Psychosis and Test of the Immune Subgroup Hypothesis.''
\emph{Schizophrenia Bulletin} 45 (5): 1120--33.

\leavevmode\hypertarget{ref-ploderl2019chances}{}%
Plöderl, Martin, and Michael Pascal Hengartner. 2019. ``What Are the
Chances for Personalised Treatment with Antidepressants? Detection of
Patient-by-Treatment Interaction with a Variance Ratio Meta-Analysis.''
\emph{BMJ Open} 9 (12).

\leavevmode\hypertarget{ref-raudenbush1987examining}{}%
Raudenbush, Stephen W, and Anthony S Bryk. 1987. ``Examining Correlates
of Diversity.'' \emph{Journal of Educational Statistics} 12 (3):
241--69.

\leavevmode\hypertarget{ref-rogdaki2020magnitude}{}%
Rogdaki, Maria, Maria Gudbrandsen, Robert A McCutcheon, Charlotte E
Blackmore, Stefan Brugger, Christine Ecker, Michael C Craig, Eileen
Daly, Declan GM Murphy, and Oliver Howes. 2020. ``Magnitude and
Heterogeneity of Brain Structural Abnormalities in 22q11. 2 Deletion
Syndrome: A Meta-Analysis.'' \emph{Molecular Psychiatry}, 1--14.

\leavevmode\hypertarget{ref-senior2016meta}{}%
Senior, Alistair M, Alison K Gosby, Jing Lu, Stephen J Simpson, and
David Raubenheimer. 2016. ``Meta-Analysis of Variance: An Illustration
Comparing the Effects of Two Dietary Interventions on Variability in
Weight.'' \emph{Evolution, Medicine, and Public Health} 2016 (1):
244--55.

\leavevmode\hypertarget{ref-spiegelhalter2019art}{}%
Spiegelhalter, David. 2019. \emph{The Art of Statistics: Learning from
Data}. Penguin UK.

\leavevmode\hypertarget{ref-volkmann2020treatment}{}%
Volkmann, Constantin Michael Dimitri, Alexander Volkmann, and Christian
Mueller. 2020. ``On the Treatment Effect Heterogeneity of
Antidepressants in Major Depression. A Bayesian Meta-Analysis.''
\emph{medRxiv}.

\leavevmode\hypertarget{ref-winkelbeiner2019evaluation}{}%
Winkelbeiner, Stephanie, Stefan Leucht, John M Kane, and Philipp Homan.
2019. ``Evaluation of Differences in Individual Treatment Response in
Schizophrenia Spectrum Disorders: A Meta-Analysis.'' \emph{JAMA
Psychiatry} 76 (10): 1063--73.

\end{document}